\documentclass[a4paper,12pt]{article}
\input epsf
\title{\vskip 30pt
\textsc{Vortex Filament in Three-manifold \\
and the Duistermaat-Heckman Formula}
}
\author{\bf{Yukinori Yasui}
\thanks{E-mail: h1879@ocugw.cc.osaka-cu.ac.jp}\\
\it{Department of Physics, Osaka City University,}\\
\it{Sugimoto, Sumiyoshi-Ku, Osaka 558, Japan}\\
   \and \bf{Waichi Ogura}
\thanks{E-mail: ogura@funpth.phys.sci.osaka-u.ac.jp}\\
\it{Department of Physic, Osaka University,}\\
\it{Machikaneyama, Toyonaka-Shi, Osaka 560, Japan}}
\newcommand{\ip}[2]{\ensuremath{\langle #1 , #2 \rangle}}
\newcommand{\pd}[2]{\ensuremath{\frac{\partial #1}{\partial #2}}}
\newcommand{\dpd}[2]{\ensuremath{\frac{\partial^2 #1}{\partial
#2^2}}}
\newcommand{\norm}[1]{\ensuremath{\parallel #1 \parallel}}
\newcommand{\half}{\ensuremath{\frac{1}{2}}}
\newcommand{\grad}{\ensuremath{\mathrm{grad}\,}}
\newcommand{\Ric}{\ensuremath{\mathrm{Ric}\,}}
\newcommand{\T}{\ensuremath{\mathsf T}}
\newcommand{\N}{\ensuremath{\mathsf N}}
\newcommand{\B}{\ensuremath{\mathsf B}}
\newcommand{\R}{\ensuremath{\mathsf R}}
\newcommand{\C}{\ensuremath{\mathsf C}}
\newcommand{\Z}{\ensuremath{\mathsf Z}}
\begin{document}
\maketitle
\vskip -125mm
\rightline{
\begin{tabular}{l}
OU-HET 219\\
hep-th/9508118
\end{tabular}}
\vskip 125mm
\vskip 60pt
\begin{abstract}
Symplectic geometry of the vortex filament in a curved
three-manifold is investigated.
There appears an infinite sequence of constants of motion in
involution in the case of constant curvature.
The Duistermaat-Heckman formula is examined perturbatively for the
classical partition function in our model and verified up to the
3-loop order.
\end{abstract}

\pagebreak
\parindent=24pt
\parskip=0pt
\baselineskip=18pt

\section{Introduction}
Kinematics of a very thin vortex tube in three-dimensional fluid may
be described by the filament equation in the local induction
approximation\ \cite{LI1,LI2}.
It is formulated as
\begin{equation}
\pd{\gamma}{t}=\pd{\gamma}{s}\times\dpd{\gamma}{s},
\label{filament}
\end{equation}
where $\gamma=\gamma\,(t,s)$ denotes the position of the vortex
filament in $\R^3$ with $t$ and $s$ being the time and the
arc-length parameter respectively.

Hasimoto\ \cite{H} introduced a map $h:\gamma\mapsto\psi=\kappa\,
\exp [i\int^s\tau(u)du]$, in order to transform the filament
equation into the nonlinear Schr\"{o}dinger (NLS) equation for
$\psi$.
Here $\kappa$ and $\tau$ respectively denote the curvature and the
torsion along $\gamma$.
Since the integrability of the NLS equation was well known, the
filament equation was naturally expected to be integrable.
Mardson and Weinstein\ \cite{MW} first described the filament
equation as a Hamiltonian equation with the Hamiltonian simply
being the length $\ell\,[\gamma]$ of the vortex filament.
Later Langer and Perline\ \cite{LP} used this Hamiltonian structure
to prove the existence of an infinite sequence of constants of
motion in involution, and studied the evolution of the vortex
filaments in connection with the solitons in the NLS equation.

With this concern in mind, we have investigated the filament
equation in a curved three-manifold $M$.
Although Langer and Perline have limited $M$ to $\R^3$, we find an
analogous integrable hierarchy in the case of constant curvature.
We further study the classical partition function for the vortex
filaments
\begin{equation}
Z(\beta)=\int_{\Gamma}\, e^{-\beta\,\ell\,[\gamma]}\,{\cal D}
\gamma{}.
\label{partition}
\end{equation}
It is not clear if the Duistermaat-Heckman formula\ \cite{DH}
applies to this case, because our phase space $\Gamma$ is neither
finite dimensional nor compact, and furthermore because the
Hamiltonian flow may not be periodic.
But the perturbative calculation in our mode reveals that the loop
corrections to the formula vanish up to the 3-loop.

\section{Integrability}

We begin this section by describing a symplectic structure for the
vortex filament in a three-manifold $M$ equipped with a Riemann
metric $g$.
Everything is considered in the smooth category for simplicity.
Let $\Gamma$ be the space of vortex filaments with fixed end points
$p$ and $q$;
$\Gamma$ is the quotient space of $\{\gamma:[0,1]\rightarrow M
\mid\gamma(0)=p,\gamma(1)=q\}$ with the reparametrization of
$\gamma$.
Hereafter $\gamma$ denotes the representative for which
the parameter $x\in[0,1]$ is a multiple of the arc-length $s$,
namely
\begin{equation}
\frac{ds}{dx}=\norm{\frac{d\gamma}{dx}}=
\sqrt{(\frac{d\gamma}{dx},\frac{d\gamma}{dx})}
\end{equation}
is independent of $x$.
Here $(\ \,,\ )$ denotes the inner product on
the tangent space $T_{\gamma\,(x)}M$.
One can identify the tangent space $T_{\gamma}\Gamma$ with the
subspace of $\Gamma(\gamma^{\ast}TM)$, and expand
$X\in\Gamma(\gamma^{\ast}TM)$
in the Frenet-Serret frame along $\gamma$ such that
\begin{equation}
X=f\,\T+g\,\N+h\,\B,
\end{equation}
where $\T$ is the unit tangent vector to $\gamma$, $\N$ is the unit
normal vector and $\B$ is the unit binormal vector.
Let $\ell\,[\gamma]$ be the length of $\gamma$, so that
$s=\ell\,[\gamma]\,x$. The Frenet-Serret equations are
\begin{equation}
\nabla_{s}\T=\kappa\,\N,\ \ \ \
\nabla_{s}\N=-\kappa\,\T+\tau\,\B,\ \ \ \
\nabla_{s}\B=-\tau\,\N,
\label{FS}
\end{equation}
with $\nabla$ being the connection on  $\gamma^{\ast}TM$ induced by
the Levi-Civita connection on $TM$.
Let $\wp$ be the projection from $\Gamma(\gamma^{\ast}TM)$ to
$T_{\gamma}\Gamma$,
then one can show that the tangent component of $v=\wp\,(X)\in
T_{\gamma}\Gamma$ satisfies
\begin{equation}
\frac{d}{dx}\,v_{\T}=\ell^{-1}(\nabla_{x}v,\frac{d\gamma}{dx})+
\ell\,\kappa\,v_{\N},
\end{equation}
and $(\nabla_{x}v,d\gamma/dx)$ is a constant.
Fixing this constant by the boundary conditions $X(0)=X(1)=0$,
one obtains
\begin{equation}
\wp\,(X)=v=\ell\,(\int^{x}_{0}\kappa\,v_{\N}dx -
x\,\int^{1}_{0}\kappa\,v_{\N}dx)\,\T +v_{\N}\N+v_{\B}\B\ .
\end{equation}

Geometrical structures on $\Gamma$ were first studied by Marsden
and Weinstein\ \cite{MW} for the vortex filament in $\R^3$,
and generalized to the loop space for a three-manifold $M$ by
Brylinski\ \cite{B}.
It is straightforward to find those for the vortex filament in $M$.

{\parindent=0pt i) Complex structure}

For the tangent vector $v\in T_{\gamma}\Gamma$, $J$ generates the
90-degree rotation
\begin{equation}
J(v)=-\,\wp\,(\T\times v)\ ,\ \ \ J^2=-1.
\end{equation}
Choosing $(v_{\N},v_{\B})$ as coordinates for $T_{\gamma}\Gamma$,
we find that $J$ corresponds to the multiplication by $i$ for the
complex function $v_{\N}(x)+i\,v_{\B}(x)$.
Hence $J$ induces a complex structure on $\Gamma$.

{\parindent=0pt ii) Riemann structure}

The Riemann structure on $\Gamma$ is simply defined by
\begin{equation}
\ip{u}{v}_{\Gamma}=\ell\,\int_{0}^{1}(u_{\N}v_{\N}+u_{\B}v_{\B})\,dx
\end{equation}
for $u,v\in T_{\gamma}\Gamma$, and satisfies the hermitian condition
\begin{equation}
\ip{u}{v}_{\Gamma}=\ip{J(u)}{J(v)}_{\Gamma}.
\end{equation}
Note that even though $\ip{\ \,}{\ }_{\Gamma}$ ignores the
$\T$-components, it is non-degenerate.

{\parindent=0pt iii) Symplectic structure}

The volume form $\nu$ on $M$ associated with the Riemann metric
$g$ provides the symplectic structure on $\Gamma$, namely
\begin{equation}
\omega(u,v)=\int_0^1\nu(\frac{d\gamma}{dx},u,v)\,dx.
\end{equation}
Using the Frenet-Serret frame, one can rewrite this as
\begin{equation}
\omega(u,v)=\ell\,\int_0^1(u_{\N}v_{\B}-u_{\B}v_{\N})\,dx,
\end{equation}
which is equivalent to the one constructed from the above two
structures
\begin{equation}
\omega(u,v)=\ip{u}{J(v)}_{\Gamma}.
\end{equation}

Having set out the basic structures, we now turn to the Hamiltonian
flows for the vortex filament.
Let $\ell\,:\Gamma\mapsto\R$ be a smooth Hamiltonian function,
then the Hamiltonian vector field $X_{\ell}$ has the form
\begin{equation}
X_{\ell}=J(\grad\ell).
\end{equation}
Choosing $i_{X_{\ell}}\,\omega=d\,\ell$ and putting
$v=d\gamma_t/dt\mid_{t=0}$, we get
\begin{eqnarray}
v\,\ell\,[\gamma]&=&\frac{d}{dt}\left.\int_0^1
\sqrt{\left(\pd{\gamma_t}{x},
\pd{\gamma_t}{x}\right)}dx\right|_{t=0},\nonumber\\
&=&\frac{1}{\ell\,[\gamma]}\int_0^1
\left(\nabla_xv,\frac{d\gamma}{dx}\right)dx,\\
&=&-\ell\,[\gamma]\,\int_0^1\left(v,\kappa\,\N\right)dx.\nonumber
\end{eqnarray}
$\grad\ell=-\wp(\kappa\,\N)$ follows, and therefore
\begin{equation}
X_{\ell}=\kappa\,\B.
\end{equation}
This yields a natural generalization of the filament equation in
$M$\ \cite{K}
\begin{equation}
\pd{\gamma}{t}=\kappa\,\B
=\ell^{-3}\,\pd{\gamma}{x}\times\nabla_{x}\pd{\gamma}{x}.
\end{equation}
The evolution equations for $\kappa$ and $\tau$ are the followings
\begin{eqnarray}
\pd{\kappa}{t}&=&\kappa\,\Ric(\B,\N)-\ell^{-1}
(2\tau\pd{\kappa}{x}+\kappa\pd{\tau}{x}),\\
\pd{\tau}{t}&=&\tau\,\Ric(\T,\N)+\ell^{-1}
\pd{}{x}(\frac{1}{2}\kappa^2+\ell^{-2}\kappa^{-1}\dpd{\kappa}{x}
-\tau^{2}+\rho(\T,\B)),
\end{eqnarray}
where $\Ric$ and $\rho$ denote the Ricci tensor and the sectional
curvature on $M$ respectively.
In the case of constant curvature, these equations take
simpler forms
\begin{eqnarray}
\pd{\kappa}{t}&=&-\ell^{-1} (2\tau\pd{\kappa}{x}+
\kappa\pd{\tau}{x}),\label{evol1}\\
\pd{\tau}{t}&=&\ell^{-1}\pd{}{x}
(\frac{1}{2}\kappa^2+\ell^{-2}\kappa^{-1}\dpd{\kappa}{x}
-\tau^{2}),\label{evol2}
\end{eqnarray}
which turn out to be identical with the equations appeared
in\ \cite{H}.
Hence we can get the following proposition for a three-manifold
with constant curvature.
\vskip 8pt
{\parindent=0pt {\bf Proposition}}
\begin{itemize}
\item[(a)] The filament equation is transformed into the NLS
equation by the Hasimoto map.
\item[(b)] There is an infinite sequence of constants of motion.
\item[(c)] These constants are in involution.
\end{itemize}
\vskip 8pt

{\parindent=0pt {\bf Proof}}

We assume that $\kappa$, $\tau$ and their derivatives of arbitrary
order vanish at the boundaries.
Then it is straightforward to prove (a) and (b) due to the evolution
equations\ (\ref{evol1}) and (\ref{evol2}).
Using the explicit form of the Hamiltonian vector fields $X_n$ (see
Remark (2)), we can confirm the commutativity $\omega(X_n,X_m)=0$
for any $n$ and $m$ with the help of\ \cite{LP}, and consequently
prove (c).
\vskip 8pt

{\parindent=0pt {\bf Remarks}}
\begin{itemize}
\item[(1)] The constants of motion are as follows\ \cite{LP}:
\begin{eqnarray}
I_{-2}[\gamma]&=&\ell\,[\gamma], \ \ \ I_{-1}[\gamma]=
\ell\,\int_{0}^{1}\tau\,dx,\\
I_{n}[\gamma]&=&\tilde I_{n}\circ h[\gamma]\ \ \ (n=0,1,2,\ldots),
\nonumber
\end{eqnarray}
where $h$ is the Hasimoto map
$h[\gamma]=\kappa\,\exp[i\ell\,[\gamma]\int_0^x\tau\,dx]$, and
$\tilde I_{n}$'s are the constants of motion in the NLS
equation\ \cite{FT} given by
\begin{equation}
\tilde I_{n}[\psi]=\ell\,\int_{0}^{1}
\half\,\bar\psi\,\tilde J_{n}(\psi,\bar\psi)\,dx,
\end{equation}
and
\begin{equation}
\tilde J_{0}=\psi,\ \ \
\tilde J_{n+1}=-i\,\frac{d}{ds}\tilde J_{n}-
\frac{1}{4}\,\bar\psi\sum_{k=1}^{n}\tilde J_{k-1}\,\tilde J_{n-k}.
\end{equation}
\item[(2)] We find mutually commuting Hamiltonian vector fields
$X_{n}$ for $I_{n}[\gamma]$:
\begin{eqnarray}
X_{-2}&=&\kappa\,\B, \ \ \ X_{-1}=R\,X_{-2},\\
X_{n}&=&R^{n+2}X_{-2}-c\,R^{n}X_{-2}\ \ \ (n=0,1,2,\ldots),
\nonumber
\end{eqnarray}
where $c$ denotes the constant curvature and $R$ the ``recursion
operator" defined by
\begin{equation}
R(v)=-\ell^{-1}\,\wp(\T\times\nabla_{x}v)
\end{equation}
for $v\in T_{\gamma}\Gamma$.
$R$ coincides with the one appeared in\ \cite{LP} when we restrict
$v$ to the Hamiltonian vector fields.
\item[(3)] Langer and Perline interpreted the Hasimoto map as a
Poisson map between the Poisson structure on the space of the
vertex filaments
and the ``forth" Poisson structure on the space of the NLS fields.
We have found no such correspondence in our model, because the
deformation of the vortex filament also changes its length $\ell$.
In the case of\ \cite{LP}, however, the vortex filament extends
boundlessly, so that the arc-length parameter is simply a parameter
and does not change under the deformation.
A different approach to the integrability of the vortex filaments
has been investigated in\ \cite{S} recently.
\end{itemize}
The filament equation belongs to an infinite hierarchy of
Hamiltonian systems
$\{\partial\gamma/\partial t_n=X_n\mid n=-2,-1,0,\ldots\}$,
and all Hamiltonian flows in this hierarchy are transformed into
those in the NLS hierarchy.
In fact, the differential of $h$ yields
\begin{equation}
dh:X_n\longmapsto\tilde{X}_{n+4}-2c\,\tilde{X}_{n+2}+c^2\,
\tilde{X}_{n}\ \ \ \pmod{i\psi},
\end{equation}
where $\tilde{X}_{-2}=\tilde{X}_{-1}=0$, and $\tilde{X}_{n}\
(n=0,1,2,\ldots)$
are the Hamiltonian vector fields associated with
$\tilde{I}_n[\psi]$,
{\it i.e.}, $\tilde{X}_{n}=-i\,\grad\tilde{I}_{n}$; first two are
\begin{eqnarray}
dh(X_{-2})&=&i\,(\frac{d^2\psi}{ds^2}+
\frac{1}{2}\mid\psi\mid^{2}\psi),\\
dh(X_{-1})&=&\frac{d^3\psi}{ds^3}+
\frac{3}{2}\mid\psi\mid^{2}\frac{d\psi}{ds}+
2c\,\frac{d\psi}{ds}.
\end{eqnarray}

\section{Classical partition function}

In this section we evaluate the classical partition function\
(\ref{partition}) with ${\cal D}\gamma$ being the symplectic volume
form on $\Gamma$.
The stationary phase method provides an asymptotic expansion for
$Z(\beta)$ as $\beta\mapsto\infty$, such that
\begin{equation}
Z(\beta)=\sum_{\grad\ell\,[\gamma]\,=0}Z_{\mathrm
WKB}[\gamma,\beta]\,
(1+\frac{a_1[\gamma]}{\beta}+\frac{a_2[\gamma]}{\beta^2}+\cdots).
\label{asymptotic}
\end{equation}
The exactness of the stationary phase (WKB) approximation has been
of interest due to the Duistermaat-Heckman formula\ \cite{DH},
where they have shown that if $\Gamma$ is a compact symplectic
manifold and $\ell$ is a periodic Hamiltonian with isolated
critical points,
WKB approximation becomes exact for\ (\ref{partition}), {\it i.e.},
the asymptotic expansion terminates at $Z_{\mathrm WKB}$.
In more general arguments presented in\ \cite{A}, the fixed points
are not necessarily isolated, and it is not mandatory to consider
the circle action alone according to the analogous results obtained
for higher dimensional tori.
For the infinite dimensional symplectic manifolds, the WKB exactness
has not been proved rigorously, but a ``proper" version of WKB
approximation should yield a reliable result for a large class of
integrable models\ \cite{N1,N2,N3,N4}.
With this notion in mind, we present the explicit calculation of the
asymptotic expansion\ (\ref{asymptotic}).
For simplicity, we will assume the followings:
\begin{itemize}
\item[(1)] $M$ is a three-manifold with a constant curvature $c$,
so that the filament equation is integrable in the sense of
Proposition.
\item[(2)] Two points $p$ and $q$ on $M$ are not conjugate.
Consequently, the Hamiltonian $\ell$ is a Morse function on
$\Gamma$,
{\it i.e.}, critical points are the geodesics on $M$ connecting
$p$ and $q$,
and further the Hessian operator $H_{\gamma}$ at each geodesic
$\gamma$ is a non-degenerate Jacobi operator
\begin{equation}
H_{\gamma}=-\nabla_{x}\nabla_{x}-c\,\ell[\gamma]^2.
\end{equation}
\end{itemize}
Let us first expand the Hamiltonian $\ell$ around a geodesic
$\gamma$.
As we can see in (\ref{FS}), the curvature along the geodesic
vanishes identically, and $\xi(x)\in T_{\gamma(x)}M$ thus
satisfies the condition $(\xi,\T)=0$.
Using an infinitesimal deformation of $\gamma$ generated by the
exponential map
$\gamma_s(x)=\exp_{\gamma(x)}[s\,\ell\,\xi(x)]$,
we can find the expansion
\begin{eqnarray}
\ell\,[\gamma_s]&=&\sum_{n=0}^{\infty}\frac{s^n}{n!}
\left.\frac{d^n}{ds^n}\ell\,[\gamma_s]\right|_{s=0},\\
&=&\ell\,[\gamma]\sum_{n=0}^{\infty}\frac{s^{2n}}{(2n)!}
\int_0^1 W_{2n}(\xi)dx.
\end{eqnarray}
Here the integrand $W_{2n}$ is given by the Bell Polynomial
$Y_m$\ \cite{Bell}, namely
\begin{equation}
W_{2n}(\xi)=Y_{2n}(f_{1},\ldots,f_{2n};g_1(\xi),\ldots,g_{2n}(\xi)),
\end{equation}
with
\begin{eqnarray}
f_m&=&(-)^{m-1}\frac{(2m-3)!!}{2^m},\nonumber\\
g_2(\xi)&=&2\,(\nabla_x\xi,\nabla_x\xi)-2c\,\ell^2(\xi,\xi),
\nonumber\\
g_{2m}(\xi)&=&(-)^{m}2^{2m-1}\,\{
(c\,\ell^2)^{m}(\xi,\xi)^{m}\label{vertices}\\
&+&(c\,\ell^2)^{m-1}(\xi,\xi)^{m-2}
[(\nabla_x\xi,\xi)^2-(\xi,\xi)(\nabla_x\xi,\nabla_x\xi)]\}
\ \ \ (m\geq 2),\nonumber\\
g_{2m+1}(\xi)&=&0.\nonumber
\end{eqnarray}
First few are given by
\begin{equation}
\begin{array}{lcl}
W_0=1, & \ \ \ \ & W_2=f_1\,g_2,\\
&&\\
W_4=f_1\,g_4+3\,f_2\,g_2^2, & \ \ \ \ &
W_6=f_1\,g_6+15\,f_2\,g_4\,g_2+15\,f_3\,g_2^3.\\
\end{array}
\end{equation}
Now let us evaluate the WKB partition function
\begin{eqnarray}
Z_{\mathrm WKB}[\gamma,\beta]&=&e^{-\beta\ell[\gamma]}
\int{\cal D}\,\xi
\exp\left[-\frac{\beta\ell[\gamma]}{2}\int_0^1W_2(\xi)dx\right],
\label{WKB}\\
\int_0^1W_2(\xi)dx&=&\ip{\xi}{H_{\gamma}(\xi)}_{\Gamma}.
\end{eqnarray}
Using the zeta-function regularization technique, we can perform the
infinite dimensional integral in\ (\ref{WKB}), and obtain\ \cite{AS}
\begin{equation}
Z_{\mathrm WKB}[\gamma,\beta]=
e^{-\beta\,\ell[\gamma]\pm\frac{\pi}{4}
i(\eta_{H}(0)-\zeta_{H}(0))}
e^{\frac{1}{2}\zeta_{H}'(0)}\,
(\beta\,\ell[\gamma])^{-\frac{1}{2}\zeta_{H}(0)},
\end{equation}
with $\eta_{H}(z)$ and $\zeta_{H}(z)\ (z\in\C)$ being eta and zeta
functions associated with the Hessian operator $H_{\gamma}$
respectively.
Evaluating these functions for $\gamma$ with the Morse index
$\mu(\gamma)$, we find
\begin{equation}
\eta_{H}(0)=-1-2\,\mu(\gamma), \ \ \ \zeta_{H}(0)=-1, \ \ \
\zeta_{H}'(0)=2\ln\left|\frac{\sqrt{c}\,\ell[\gamma]}
{2\sin(\sqrt{c}\,\ell[\gamma])}\right|,
\end{equation}
and eventually this gives us an explicit expression
\begin{equation}
Z_{\mathrm WKB}[\gamma,\beta]=\frac{1}{2}\,e^{-\beta\,\ell[\gamma]}
\sqrt{\beta\,\ell[\gamma]}
\left|\frac{\sqrt{c}\,\ell[\gamma]}{\sin(\sqrt{c}\,\ell[\gamma])}
\right|
e^{\mp\frac{\pi}{2}i\,\mu(\gamma)}.
\end{equation}
Since $\mu(\gamma)$ is an even integer, the last factor contains no
ambiguities.

We now proceed to the higher-order calculation.
It is convenient to choose an orthogonal frame $\{e_1,e_2\}$ along
$\gamma$ such that
\begin{equation}
\nabla_x\,e_i=0,\ \ \ (\T,e_i)=0 \ \ \ \mathrm {for} \ i=1,2.
\end{equation}
In this frame, the kernel of the Jacobi operator $H_{\gamma}$
becomes diagonal,
and both of the diagonal elements are identical to the Dirichlet
Green function
\begin{equation}
G(x,x')=2\sum_{n=1}^{\infty}\frac{\sin(n\pi x)\sin(n\pi x')}
{(n\pi)^2-\lambda},
\end{equation}
with $\lambda=c\,\ell^2$.
The 2-loop amplitude $a_1=-\beta^2\,\langle W_4/4!\rangle$ consists
of four diagrams depicted in Fig.\ 1, and those are respectively
\begin{eqnarray}
&(a)\hskip 30pt& \lambda^2\,\int_0^1\,G(x)^2=
\frac{\lambda}{8}-\frac{3}{8}\sqrt{\lambda}\,X+\frac{3}{8}\lambda\,
X^2, \nonumber\\
&(b)\hskip 30pt& \lambda\,\int_0^1\,G'(x)^2=
\frac{\lambda}{8}-\frac{1}{8}\sqrt{\lambda}\,X+\frac{1}{8}\lambda\,
X^2, \nonumber\\
&(c)\hskip 30pt& \lambda\,\int_0^1\,G(x)\,G''(x)=
-\frac{\lambda}{8}-\frac{1}{8}\sqrt{\lambda}\,X+\frac{1}{8}\lambda\,
X^2,\nonumber\\
&(d)\hskip 30pt& \,\int_0^1\,G''(x)^2=\lambda^2\,\int_0^1\,G(x)^2,
\nonumber
\end{eqnarray}
where $X=\cot\sqrt{\lambda}$,
$G(x)=G(x,x')\mid_{x=x'}$,
$G'(x)=(\partial/\partial x)\,G(x,x')\mid_{x=x'}$ and
$G''(x)=(\partial^2/\partial x\,\partial x')\,G(x,x')\mid_{x=x'}$.
While $G(x)$ and $G'(x)$ are convergent, $G''(x)$ diverges at the
boundaries, thus we have found (c) and (d) by executing the
$x$-integration first and then by regularizing the infinite
$n$-summation in terms of the following analytic continuations:
\begin{eqnarray}
\sum_{n=1}^{\infty}\frac{1}{(n^2+a^2)^s}&=&
-\half a^{-2s}+\frac{\pi^{\half}}{2}\,
\frac{\Gamma(s-\half)}{\Gamma(s)}\,a^{-2s+1}\nonumber\\
&+&2\,\frac{\pi^{\half}}{\Gamma(s)}\sum_{n=1}^{\infty}
\left(\frac{\pi n}{a}\right)^{s-\half}\,K_{s-\half}(2\pi na),\\
\sum_{n=1}^{\infty}\frac{n^2}{(n^2+a^2)^s}&=&
\frac{\pi^{\half}}{4}\,\frac{\Gamma(s-\frac{3}{2})}{\Gamma(s)}\,
a^{-2s+3}\nonumber\\
&+&\frac{\pi^{\half}}{\Gamma(s)}\,\sum_{n=1}^{\infty}
\left(\frac{\pi n}{a}\right)^{s-\frac{3}{2}}\,
K_{s-\frac{3}{2}}(2\pi na)\\
&-&2\,\frac{\pi^{\frac{5}{2}}}{\Gamma(s)}\,\sum_{n=1}^{\infty}
\,n^2\,\left(\frac{\pi n}{a}\right)^{s-\frac{5}{2}}\,
K_{s-\frac{5}{2}}(2\pi na),\nonumber\\
\sum_{n=1}^{\infty}\frac{n^4}{(n^2+a^2)^s}&=&
-\frac{3}{8}\,\pi^{\half}\,
\frac{\Gamma(s-\frac{5}{2})}{\Gamma(s)}\,
a^{-s+\frac{5}{2}}\nonumber\\
&+&\frac{3}{2}\,\frac{\pi^{\half}}{\Gamma(s)}\,
\sum_{n=1}^{\infty}\left(\frac{\pi n}{a}\right)^{s-\frac{5}{2}}\,
K_{s-\frac{5}{2}}(2\pi na)\nonumber\\
&-&6\,\frac{\pi^{\frac{5}{2}}}{\Gamma(s)}\,\sum_{n=1}^{\infty}
\,n^2\,\left(\frac{\pi n}{a}\right)^{s-\frac{7}{2}}\,
K_{s-\frac{7}{2}}(2\pi na)\\
&+&2\,\frac{\pi^{\frac{9}{2}}}{\Gamma(s)}\,\sum_{n=1}^{\infty}
\,n^4\,\left(\frac{\pi n}{a}\right)^{s-\frac{9}{2}}\,
K_{s-\frac{9}{2}}(2\pi na),\nonumber
\end{eqnarray}
where $K_{\nu}(z)$ is the modified Bessel function.
Multiplying (a) through (d) with the weights of the diagrams,
we conclude that the 2-loop amplitude vanishes.
Beyond the 2-loop, however, we ought to generalize the analytic
continuation for a multiple infinite summation.
One might think that applying the analytic continuation method
directly to the Green function, we could regularize the Green
function, and thereby making all loop amplitude finite.
This is certainly true, but regularizing the Green function in this
way, we also eliminate the necessarily singularity at $x=x'$, and
obtain non-vanishing 2-loop amplitude as a result.
We may avoid this difficulty by treating $G(x,x')$ as a
distribution w.r.t.\ $x$.
Let us first examine this on the 2-loop and check if the amplitude
vanishes.
Since $G(x,x')$ may naturally be extended periodically (period 2)
to $\R$ as a function of $x$,
one can redefine it as a distribution $\tilde G(x,x')$ such that
\begin{eqnarray}
\tilde G(x,x')&=&-\frac{1}{\sqrt{\lambda}\sin\sqrt{\lambda}}
\sum_{n\in\Z}\,
\left\{\sin[\sqrt{\lambda}\,(x-2n)]\,\sin[\sqrt{\lambda}\,
(x'-1)]\,H(x;2n,x'+2n)\right.\nonumber\\
&\ &\hskip 30pt
+\left.\sin[\sqrt{\lambda}\,x']\,\sin[\sqrt{\lambda}\,(x-2n-1)]\,
H(x;x'+2n,2n+1)\right\}\label{Green}\\
&+&\frac{1}{\sqrt{\lambda}\sin\sqrt{\lambda}}
\sum_{n\in\Z}\,\left\{\,x\rightarrow -x\,\right\},\nonumber
\end{eqnarray}
where $H(x;a,b)$ denotes the characteristic function for the
interval $[a,b]\subset\R$.
Similarly $\tilde G(x)$ may also be extended periodically (period 1)
to $\R$
\begin{equation}
\tilde G(x)=-\frac{1}{\sqrt{\lambda}\sin\sqrt{\lambda}}
\sum_{n\in\Z}\,\left\{\sin[\sqrt{\lambda}\,(x-n)]\,
\sin[\sqrt{\lambda}\,(x-n-1)]\,H(x;n,n+1)\right\}.
\label{loop}
\end{equation}
Using the periodic delta function $\delta(x;n)$ ($n$ is the period),
we may evaluate the second derivative
\begin{eqnarray}
\frac{\partial^2}{\partial x\,\partial x'}\,\tilde G(x,x')&=&
-\frac{\sqrt{\lambda}}{\sin\sqrt{\lambda}}\sum_{n\in\Z}\,
\left\{\cos[\sqrt{\lambda}\,(x-2n)]\,\cos[\sqrt{\lambda}\,(x'-1)]\,
H(x;2n,x'+2n)\right.\nonumber\\
&\ &\hskip 30pt
+\left.\cos[\sqrt{\lambda}\,x']\,\cos[\sqrt{\lambda}\,(x-2n-1)]\,
H(x;x'+2n,2n+1)\right\}\\
&-&\!\!\!\frac{\sqrt{\lambda}}{\sin\sqrt{\lambda}}
\sum_{n\in\Z}\,\left\{\,x\rightarrow -x\,\right\}+
\delta(x-x';2)+\delta(x+x';2),\nonumber
\end{eqnarray}
and similarly
\begin{equation}
\tilde G''(x)=-\lambda\,\tilde
G(x)-\sqrt{\lambda}\,\cot\sqrt{\lambda}+
\frac{1}{2}\,\delta(x;1).
\end{equation}
The delta function appears only in $\tilde G''(x)$, and we confirm
the vanishing of the 2-loop amplitude by using
\begin{equation}
\int_0^1dx\,\delta(x;1)^2=\delta(0;1)=0,
\end{equation}
which is consistent with the
$\zeta$-function regularization because of
$\delta(0;1)=1+2\,\zeta(-1)$.

The 3-loop amplitude
$a_2=\beta^3\,\langle\beta (W_4/4!)^2/2-W_6/6!\rangle$
consists of 30 diagrams depicted in Fig.\ 2.
Evaluating them by means of\ (\ref{Green}), (\ref{loop}) and
$\delta(0;1)=0$,
we find that 29 diagrams contain no ambiguities due to the the
integration formulae
\begin{eqnarray}
\int_0^1dx\int_0^xdy\,\delta(x;1)\,\delta(y;1)\,F(x,y)
&=&\frac{1}{8}\,F(0,0)+\frac{1}{4}\,F(1,0)+\frac{1}{8}\,F(1,1),\\
\int_0^1dx\int_0^xdy\,[\delta(x-y;2)+\delta(x+y;2)]&\ &\hskip -24pt
[\delta(x;1)+\delta(y;1)]\,F(x,y)\nonumber\\
&=&\frac{1}{2}\,F(0,0)+\frac{1}{2}\,F(1,1),\label{pq}\\
\int_0^1dx\int_0^xdy\,[\delta(x-y;2)+\delta(x+y;2)]^2\,F(x,y)&=&
\frac{1}{8}\,F(0,0)+\frac{1}{8}\,F(1,1).
\end{eqnarray}
Here the last equality follows from $\delta(0;2)=0$.
Yet, in the diagram whose weight is $-480$, we encounter an
ambiguous integral
\begin{equation}
\int_0^1dx\int_0^xdy\,[\delta(x-y;2)+\delta(x+y;2)]\delta(x;1)\,
F(x,y)=p\,F(0,0)+q\,F(1,1),
\end{equation}
where $p+q=1/2$ as is shown in\ (\ref{pq}), but $p$ or $q$ alone
cannot be determined unless we specify the regularization of the
delta function.
If we were able to define the analytic continuation of the infinite
double sum, this ambiguity would not appear,
but we have no choice at our hand other than putting $q=1/16$,
and obtain the vanishing 3-loop amplitude as a result.

Ambiguities appearing in higher loops are inevitable, because they
relate to the regularization ambiguity of the integration measure
${\cal D}\gamma$, which has never been defined rigorously in the
first place.
Both methods we have presented here reveal that the degree of
ambiguity gets larger as the order of loops increases.
In the analytic continuation method, ambiguity arises from the
variety of the analytic continuation applicable to the multiple
infinite summation,
whereas in the distribution method, the delta-function integration,
particularly the finite part of the boundary contribution,
is the source of the ambiguity.
Nevertheless our lower order calculations suggest that by
regularizing ${\cal D}\gamma$ order by order,
one can eliminate all higher loop corrections, and thereby
preserving the Duistermaat-Heckman formula.
The symplectic structure has been studied thoroughly in compact
finite dimensional manifolds, but little is known for the
infinite dimensional ones, which include most of the integrable
hierarchies.
This is exactly the place where the physical interests are, and the
Duistermaat-Heckman formula would throw a new light over the
integrable hierarchies as we have caught a glimpse of it here.

\vskip 12pt

The authors would like to thank Dr.\ N.\ Sasaki for helpful
discussions.

\vfill
\pagebreak
\parindent=0pt {\bf\Large Figure captions}
\vskip 12pt
Figure 1.\ \ 2-loop diagrams.
``Dot" denotes a derivative on the Green function; for instance,
the propagator with one dot represents $G'(x)$, and the one with
two dots $G''(x)$.
The attached numbers are the weights of the diagrams.
\vskip 12pt
Figure 2.\ \ 3-loop diagrams.
Since there appear no second derivatives in\ (\ref{vertices}) and
the number of derivatives is always even, double dots on a single
propagator must go to the separate vertices and the number of the
derivatives at each vertex must be even.
One must interpret dots accordingly for the diagrams with
two vertices.
Note that for a couple of diagrams in the top group and for a couple
in the middle, though the resulting diagrams are inequivalent, this
simple rule does not tell to which vertex dots are supposed to go.
In those diagrams, dots are placed closer to the vertices to which
they are supposed to go.
\vfill
\pagebreak

\ \vskip 80pt
\begin{center}
$
\begin{array}{llrl}
\mbox{\LARGE (a)}&\hskip 24pt&
\mbox{\LARGE $+$\ 8}&
\begin{picture}(80,40)(-16,12)
\unitlength=2pt
\put(10,10){\circle{20}}
\put(30,10){\circle{20}}
\end{picture}\\
&&&\\&&&\\
\mbox{\LARGE (b)}&&
\mbox{\LARGE $+$\ 16}&
\begin{picture}(80,40)(-16,12)
\unitlength=2pt
\put(10,10){\circle{20}}
\put(10,20){\circle*{3}}
\put(30,10){\circle{20}}
\put(30,20){\circle*{3}}
\end{picture}\\
&&&\\&&&\\
\mbox{\LARGE (c)}&&
\mbox{\LARGE $+$\ 32}&
\begin{picture}(80,40)(-16,12)
\unitlength=2pt
\put(10,10){\circle{20}}
\put(30,10){\circle{20}}
\put(30,0){\circle*{3}}
\put(30,20){\circle*{3}}
\end{picture}\\
&&&\\&&&\\
\mbox{\LARGE (d)}&&
\mbox{\LARGE $-$\ 24}&
\begin{picture}(80,40)(-16,12)
\unitlength=2pt
\put(10,10){\circle{20}}
\put(10,0){\circle*{3}}
\put(10,20){\circle*{3}}
\put(30,10){\circle{20}}
\put(30,0){\circle*{3}}
\put(30,20){\circle*{3}}
\end{picture}\\
\end{array}
$
\vskip 48pt
Figure 1
\end{center}
\pagebreak
\begin{center}
\epsfbox{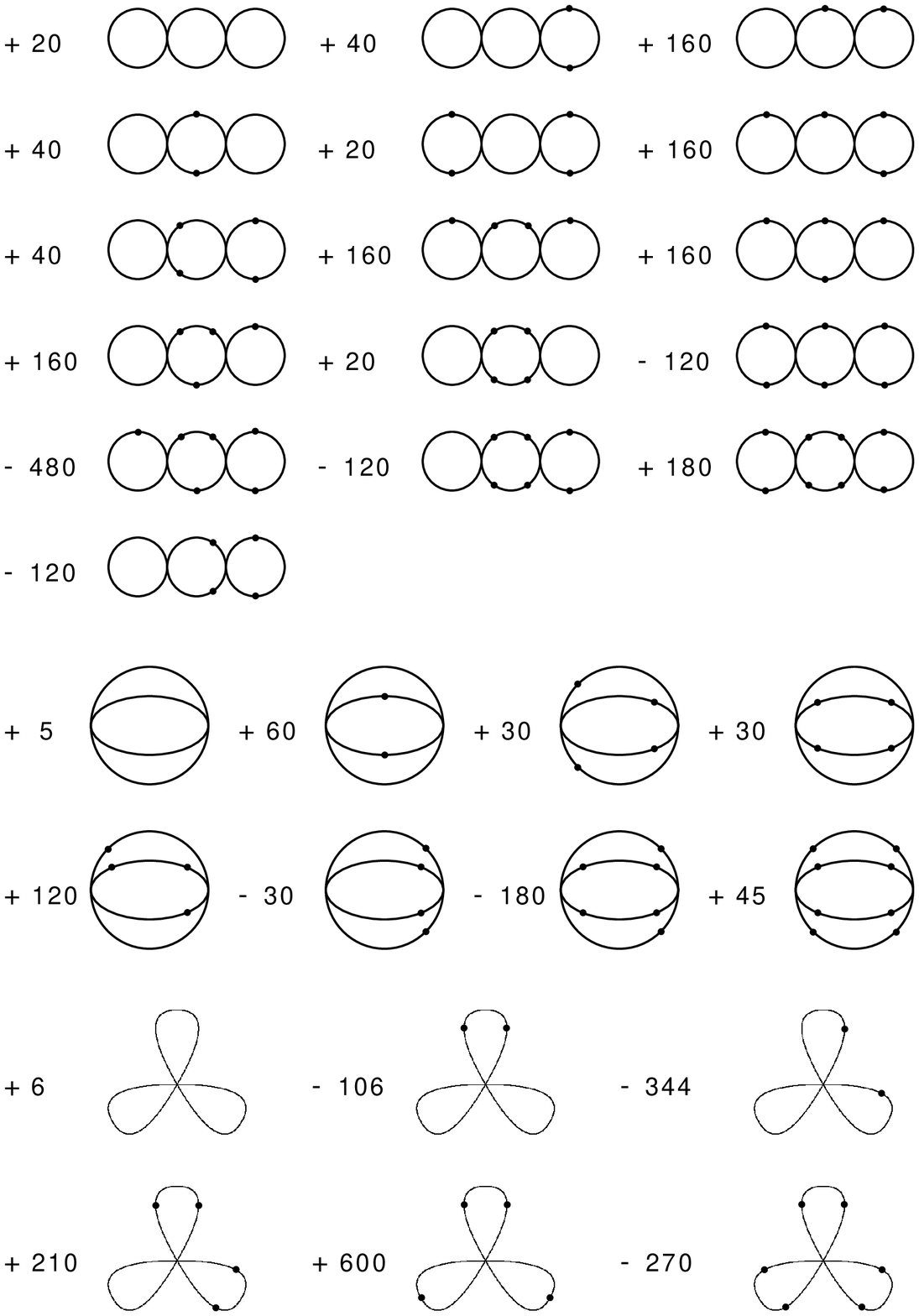}
\end{center}
\end{document}